# Comparative Analysis of Existing Methods and Algorithms for Automatic Assignment of Reviewers to Papers

Y. Kalmukov, B. Rachev

**Abstract.** The article focuses on the importance of the automatic assignment of reviewers to papers for increasing the assignment accuracy the quality of the scientific event itself. It discusses the main aspects that influence the assignment accuracy, performs a detailed analysis of the methods of describing papers and reviewers' competences used by the existing conference management systems and suggests some improvements in the way the similarity factors are calculated.

## 1. Introduction

One of the most important and challenging task in organizing scientific conferences is the assignment of reviewers to papers. Its accuracy directly impacts the quality of the conference itself. For highly-ranked conferences having a low acceptance ratio, it is crucial that each paper is evaluated by the most competent, reviewers among the Programme Committee (PC) members. Even a small inaccuracy in the assignment may cause serious and very unpleasant misjudgments that may dramatically decrease the conference image and authors' thrust in that event.

The traditional way of manually assigning reviewers to papers seems to be the most reliable way for many PC chairs, however the real-life situations prove that statement to be wrong in most cases. Manual assignment is indeed applicable for small conferences having a small number of submitted papers and reviewers well known to the PC chairs.

However as the number of papers and reviewers increases the manual assignment gets harder and harder and most importantly less and less accurate. The reason is clear – the PC chairs have to familiarize themselves with all papers and reviewers' competences, then to find somehow a way to give each paper to the most competent in its subject domain reviewer while maintaining a load balancing so that all reviewers evaluate roughly the same number of papers. Doing that, for a large number of papers and reviewers, is not just time consuming but it tends to impossible or at least improbable. That is why all modern commercially available conference management systems offer an automatic assignment of reviewers to papers and that is their key feature.

The non-intersecting sets of papers and reviewers can be represented by a complete weighted bipartite graph (*figure 1*), where P is the set of all submitted papers and R – the set of all registered reviewers. There is an edge from every paper to every reviewer and every edge has a weight. In case of a zero weight the corresponding edge may be omitted, turning the graph to a non-complete one. The weight of the edge between paper $p_i$ and reviewer $r_j$ tells us how competent (suitable) is $r_j$ to review $p_i$. This measure of suitability is called a similarity factor. The weights are calculated or assigned in accordance with the chosen way of describing papers and reviewers' competences.

The phrase „method of describing papers and competences" includes general concepts, algorithms, mathematical formulas, proper organization of the user interface etc. so that all these allow authors not just to submit a single file, but to outline what the paper is about and reviewers to state the areas of science they feel competent to review papers in. The chosen method and data model determine the exact algorithm and/or mathematical formulas to calculate the weights of the edges. Calculating a similarity factor for each pair (paper, reviewer) results in a similarity matrix (*figure 2*). It contains all the information needed for the assignment algorithm to make decisions of who will review what. In most cases the assignment algorithm takes the similarity matrix as an input and handles the assignment as an optimization problem [1] trying to find the best possible assignment.

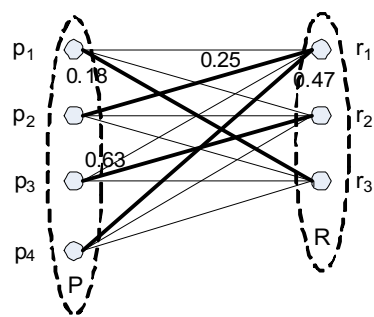

**Figure 1.** The sets of papers (P) and reviewers (R) represented as a complete weighted bipartite graph. The edges in bold are the actual assignments suggested by an assignment algorithm. All edges have weights but just those of the assignments are shown for clearness.

|       | $p_1$ | $p_2$ | $p_3$ | $p_4$ | $p_5$ |
|-------|-------|-------|-------|-------|-------|
| $r_1$ | 0.40  | 0.54  | 0.21  | 0.27  | 0.25  |
| $r_2$ | 0.20  | 0.10  | 0.14  | 0     | 0     |
| $r_3$ | 0.18  | 0.20  | 0.13  | 0     | 0     |
| $r_4$ | 0.17  | 0.30  | 0.25  | 0.17  | 0.14  |
| $r_5$ | 0.38  | 0.42  | 0.27  | 0.22  | 0.09  |

**Figure 2.** A similarity matrix containing the weights of all edges between the set of papers P and the set of reviewers R. It shows how competent is a reviewer to evaluate a specified paper.

Understanding the importance of the automatic assignment many scientists and software companies invest a large amount of efforts to improve its accuracy and efficiency. In general it depends on:

• ***The method of describing papers and competences***. It determines whether or not an additional redundant data is needed for describing papers and competences, and if users should explicitly provide it. It suggests the data model to be used for representing objects descriptions and the algorithms for calculating the similarity factors. The usage of all these results in building a similarity matrix, stating how much exactly a reviewer is competent to review a specified paper.

• ***The accuracy of the assignment algorithm itself***. The assignment algorithm takes the similarity matrix and determines the reviewers to evaluate each one of the papers.

One of the most important problems in analyzing the efficiency of the automatic assignment is how to measure the accuracy of the assignment so that two or more methods and/or algorithms could be fairly compared. As describing papers and competences involves many subjective aspects like personal judgments and personal decisions then the accuracy of two or more methods could not be fairly evaluated just by comparing (to one another) the similarity factors they produce. Instead the calculated similarities should be compared to similarity evaluations provided by real humans (PC members) over the same dataset. In contrast there is nothing subjective in the assignment algorithms as they take the same matrix of similarity factors, i.e. share the same input data, and propose an assignment the accuracy of which could be easily measured by the weight of matching.

The main goal of this article is to perform *an in-depth comparative analysis* of the automatic assignment capabilities of the most popular conference management systems. It reveals the approaches, *the methods and the algorithms* used to drive the assignment, their relevance, strengths and weaknesses in respect to the conference management. In this case papers and competences are described by a list of conference topics, the article suggests a *similarity measure*, based on Jaccard's index, which allows more precise and accurate calculation of the similarity factors in comparison with the ways the existing systems calculate them.

Since as most of the conference management systems [12,14,3,15,16,2,13] are offered as a service, not as software and their documentation is far from detailed in respect to technical issues all non-cited statements are result from the authors' own experience, assumptions and judgments. Section 2 of this article is dedicated to the methods of describing papers and competences, while section 3 deals with the algorithms for automatic assignment.

## 2. Methods of Describing Papers and Reviewers' Competences

Depending on the level of interactivity the methods of describing papers and competences can be divided into two main groups:

• Methods that require users to ***explicitly*** outline their papers and/or competences (called *Explicit methods* for short):
  ◦ Declaring an interest by *bidding*.
  ◦ Describing papers and competences by a *list of conference topics* (keywords).
  ◦ Combined – conference topics + reviewers' bidding.

• Intelligent methods that automatically extract the needed descriptive data from the papers and from the reviewers' publications available on the Internet. As they do not require any explicit user actions they are called *Implicit*

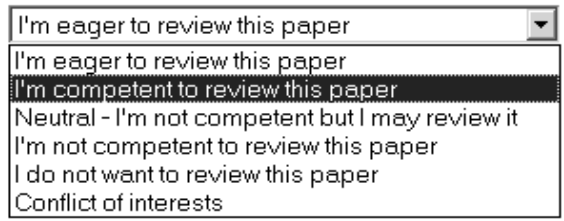

**Figure 3.** A drop down menu allowing reviewers to indicate how willing they are in reviewing a specified paper

*methods* for short. They perform a content analysis on the papers in order to automatically extract keywords or to find suitable reviewers by using the premise that the paper's subject domain is represented by the authors of its references [4].

### 2.1. Declaring Reviewers' Interest to Papers by Bidding

The main idea of *bidding*, also known as collecting *reviewers' preferences*, or *rating papers*, is that reviewers are required to browse the list of papers (abstracts) and to state their interest to each paper separately. This is usually done by selecting the relevant item from a drop down menu (*figure 3*).

An integer number corresponds to each one of the options, where the highest number is assigned to the highest level of willing to review. As the number of submitted papers is usually very high it is improbable that a reviewer will rate all of the papers. However every edge has to have a weight. So, if a paper is not rated by a specified reviewer, then the number corresponding to Neutral is assigned as weight of that edge.

**Calculating the similarity factors**

Similarity factors (weights of the edges) are explicitly assigned by reviewers by choosing the relevant interest.

**Advantages**

• The assignments made in accordance with the reviewers' preferences are considered to be 100% accurate. The reason is obvious – no one than the reviewer himself knows better if he/she is competent in the subject domain of a specified paper.

**Disadvantages**

• Although reviewers' preferences are explicitly stated, this action does not actually describe the paper neither does the reviewer. So if a paper has not been rated (chosen) by enough reviewers then the latter will be assigned to it at random, significantly decreasing the overall quality of the assignment as a whole. Here is an example: If reviewer $r_7$ indicated he is competent to review papers $p_3$ and $p_9$, there is no way for the conference management system to know if he is competent to review paper $p_{20}$ or not. If paper $p_{20}$ has not been rated by enough reviewers then reviewers will be randomly assigned to it.

**Bottom line**

The method of describing papers and competences by bidding only is the most accurate, but per „single paper" level only – i.e. if the paper is rated by enough reviewers. The accuracy of the entire assignment process however is pretty unpredictable as reviewers will never rate all of the papers in a real situation. For that reason this method is never used alone in the conference management systems, but it is usually combined with the conference topics.

### 2.2. Predicting the Missing Preferences by Using Collaborative Filtering and the Iterative Rating Method Proposed by Philippe Rigaux

The iterative rating method (IRM) [5] is a clever improvement of the bidding process which is trying to overcome the „random assignments" problem that appears when a paper has not been rated by enough reviewers. In practice the number of submitted papers is often large and it is difficult to ask for a comprehensive rating. Reviewers rate only a small subset of papers and the rating table is sparse, with many unknown values that have to be predicted in order to use an automatic assignment algorithm [5].

The method computes the predicted preferences by using a multi-step process that continuously improves the confidence level of the ratings.

Each step (iteration) consists of the following operations [5]:

1. For each reviewer – A sample of papers, whose ratings are expected to be high and to lead to the best confidence level improvement, is selected and proposed for rating;
2. Each reviewer is required to rate the papers from his/her sample;
3. Then a collaborative filtering algorithm is performed to obtain a new set of predicted ratings, based on the users ratings made so far.

At the end of each iteration the confidence level is improved and more and more reviewers' preferences are predicted. At the beginning, the first samples of papers to be rated are computed in accordance with the conference topics selected from both authors and reviewers. The main idea of the iterative rating method is illustrated on *figure 4*.

22  2 2010  information technologies and control

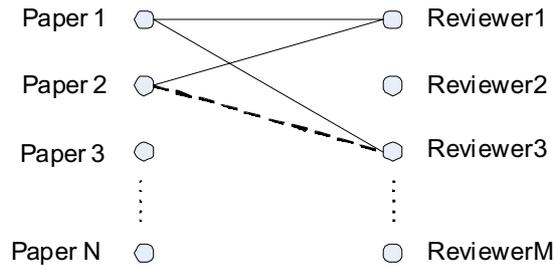

**Figure 4.** Predicting the missing ratings by using a collaborative filtering and an iterative rating method

Imagine that Reviewer 1 has explicitly stated he wants to review Paper 1 and Paper 2. Reviewer 3 has indicated he wants to review Paper 1 as well. So if both reviewers 1 and 3 have explicitly stated they want to review Paper 1, then they should have common interests and it is pretty possible that Reviewer 3 is competent to review Paper 2 as well. In this way the rating of Reviewer 3 to Paper 2 is predicted. The weight of the corresponding edge $(p_2, r_3)$ should be similar to the weight of $(p_2, r_1)$ or derivative of the weights of $(p_2, r_1)$ and $(p_1, r_3)$.

The iterative rating method is fully implemented in one of the most popular conference management systems – The MyReview System [17].

**Advantages**

• Partially (can be significantly) overcomes the „random assignments" problem by predicting many of the missing preferences.

**Disadvantages**

• Can not handle the situation where a paper has not been rated by anyone. In that case, again reviewers are assigned to it at random. In fact there is no way for this situation to be handled correctly. It is because bidding does not actually describe papers or competences, but it just adds relationships between them. If a paper has not been rated by anyone then the software will not know what the paper is about. But this is a problem of the bidding itself, not of the iterative rating method.

• Increased level of interactivity. Reviewers have to rate papers many times at multiple steps. Unfortunately, in practice, this is rarely done. In most cases the PC chairs use the iterative rating method in a single-pass mode only, not reaching its full potential.

**Bottom line**

Even when used in a single-pass mode the iterative rating method is a good improvement of the bidding process as it predicts some of the missing preferences.

## 2.3. Describing Papers and Reviewers' Competences by a List of Conference Topics (Keywords)

This method describes papers and competences independently of each other. At the beginning, before the paper submission phase, the PC chair establishes a list of conference topics (keywords or key phrases) that best describe the conference coverage area. During the paper submission authors are required to select all topics that apply to their papers. Reviewers do the same – during the registration they are required to select the topics corresponding to the areas of science they are competent in.

There are two ways, used in the existing conference management systems, of implementing this keyword mechanism:

1. By using a list of non-weighted keywords usually selected from HTML checkboxes (*figure 5*). In this way the keywords have a binary-like behavior – a keyword can be present or not in the list of keywords describing a specified paper or reviewer.

2. By using a list of user-weighted keywords usually selected from a drop down menu (*figure 6*). In this way users are able not only to state that a conference topic applies to their papers or area of expertise, but to indicate how much exactly the topic applies.

EasyChair [12] and The MyReview System [17] are the only two conference management systems that reveal in details how exactly the usage of the topics influence the automatic assignment. The official documentations of all others just state keywords influence the assignment, but not describing how. For that reasons we suggest that formulas (1), (2), (3) and (4) provide the most accurate ways of calculating the similarity factors (see below for details).

**Calculating the similarity factors in case of non-weighted keywords**

The list of keywords is in fact a binary feature vector, describing an object that may be a paper or a reviewer. In this case the Jaccard's index (1) and the Dice similarity measure (2) seem to be the most reasonable ways of calculating the similarity factors as they take into account not just the similarities between the two sets but the differences as well, i.e. show not only how competent the j-th reviewer is to evaluate the i-th paper, but also how worthy is the i-th paper to be evaluated by the j-th reviewer exactly, but not by someone else (illustrated by the following example). It does not matter whether Jaccard's or Dice's coefficient is used as the relationship between any two similarity factors calculated by using Jaccard's formula remains the same if they are calculated by Dice's measure [11].

**Figure 5.** Example of a list of non-weighted keywords. The keywords have a binary-like behavior – selected or not selected.

$$(1) \quad SFp_i r_j = w(e_{p_i r_j}) = \frac{|KWp_i \cap KWr_j|}{|KWp_i \cup KWr_j|},$$

where $SFp_i r_j$ – the similarity factor between i-th paper and j-th reviewer;
$KWp_i$ – the set of keywords, describing the i-th paper;
$KWr_j$ – the set of keywords chosen by the j-th reviewer.

$$(2) \quad SFp_i r_j = w(e_{p_i r_j}) = \frac{2 \times |KWp_i \cap KWr_j|}{|KWp_i| + |KWr_j|}.$$

Let the list of topics describing the conference coverage area is {a, b, c, d, e, f, g, h, i}.
Let a paper $p_1$ is described by topics $KWp_1$ = {b, e, g, i}, while reviewer $r_1$ has chosen $KWr_1$ = {b, c, e} and reviewer $r_2$ – $KWr_2$ = {a, b, e, f, h}. Both reviewers, $r_1$ and $r_2$, have 2 topics in common with the paper – {b, e}. If only the number of common topics matters then both reviewers are equally competent to evaluate the paper. However common sense suggests $r_1$ is more suitable to review the paper than $r_2$, first because $r_1$ is more *focused* on the paper's subject domain than $r_2$ and second - the possibility $r_2$ to be competent to review other papers as well is higher than the one of $r_1$ (because $r_2$ has chosen more keywords than $r_1$). The Jaccard's index takes both considerations into account and for that reason it is the most relevant way of calculating similarity factors in case of non-weighted topics. For this particular example $SFp_1 r_1$ = 2/5 = 0.4 while $SFp_1 r_2$ = 2/7 = 0.29, so $r_1$ is indeed a better choice according to both common sense and Jaccard's index.

**Calculating the similarity factors in case of user-weighted keywords**

In this case as the keywords are not just present or missing the formula should take into account the *exact*

**Figure 6.** Example of a list of user-weighted keywords. If a keyword is selected the author (reviewer) can state how much it applies to his/her paper (competences). „Not applicable" means not selected.

*amount of presence* of each common keyword.

Let us assume that in case of a common keyword:
* If the level of expertise (on that particular topic) of the reviewer is higher than or equal to the level chosen by the author then the reviewer is considered to be 100% competent to review the specified paper in respect to that particular topic only.
* If the level of expertise (on that particular topic) of the reviewer is less than the level chosen by the author, then the reviewer is less than 100% competent to evaluate the paper.

Such a comparison should be made for every topic in common and then the partial results should be combined in a single one. In terms of mathematics these two assumptions are represented by the system of equations (3). It is a modified version of (1), suggested by the authors of this article.

$$(3) \quad SFp_i r_j = w(e_{p_i r_j}) = \frac{\sum\limits_{kw_c = KWp_i \cap KWr_j} A_c}{|KWp_i \cup KWr_j|}$$

$$\text{If } w_{r_j}(kw_c) >= w_{p_i}(kw_c), A_c = 1$$
$$\text{else } A_c = 1 - (w_{p_i}(kw_c) - w_{r_j}(kw_c))$$

where $SFp_i r_j$ – the similarity factor between i-th paper and j-th reviewer;
$KWp_i$ – the set of keywords, describing the i-th paper;
$KWr_j$ – the set of keywords chosen by the j-th reviewer;
$kw_c$ – a common keyword describing both paper $p_i$ and reviewer $r_j$;
$w_{r_j}(kw_c)$ – the expertise of $r_j$ on the topic $kw_c$. The weight of $kw_c$ assigned by $r_j$. $w_{r_j}(kw_c) \in [0, 1]$;
$w_{p_i}(kw_c)$ – shows how much $kw_c$ applies to $p_i$. The weight of $kw_c$ assigned to $p_i$. $w_{p_i}(kw_c) \in [0, 1]$;
$A_c$ – shows how much $r_j$ is competent to review $p_i$ in respect to $kw_c$ only.

**Here is an example.**

As in the above example the list of topics describing the conference is {a, b, c, d, e, f, g, h, i}.

Paper $p_1$ is described by $KWp_1 = \{b, e, g, i\}$, while reviewers $r_1$ and $r_2$ have chosen the same topics, i.e. $KWr_1 = KWr_2 = \{b, c, e\}$. Formula (3) in contrast with (1) takes into account the exact presence of the topics $b$ and $e$ in the paper and in the reviewers' competences as well.

Let us have the following levels of competence:
$w_{r_1}(b) = 0.7$ (equals to 70%)
$w_{r_1}(e) = 0.4$         $w_{p_1}(b) = 0.5$
$w_{r_2}(b) = 0.3$         $w_{p_1}(e) = 0.4$
$w_{r_2}(e) = 0.6$

Then $SFp_1 r_1 = (1+1) / 5 = 0.4$, because $w_{r_1}(b) > w_{p_1}(b)$ and $w_{r_1}(e) = w_{p_1}(e)$.

For $SFp_1 r_2$, $w_{r_2}(b) < w_{p_1}(b)$ and $w_{r_2}(e) > w_{p_1}(e)$, so
$SFp_1 r_2 = ((1 - (0.5 - 0.3)) + 1) / 5 = 1.8 / 5 = 0.36$

Reviewer $r_1$ is more competent to review $p_1$ than $r_2$ although they both have chosen the same topics because the level of competence of $r_1$ in respect to $b$ is higher than the level $b$ applied to $p_1$, while the level of competence of $r_2$ in respect to $b$ is lower than the level $b$ applied to $p_1$.

\* Equation (3) takes into account the *relative level of competence* only, i.e. whether the reviewer is more (or less) competent in specific areas than the level they apply to the paper. In relation to the example above if there are two reviewers – $r_1$ – 70% competent in $b$, $r_2$ – 90% competent in the same topic and it applies 50% to the paper, then these two reviewers are assumed to be equally competent to review the paper in respect to that specific topic.

If the *absolute level of competence* of the reviewer in respect to a specified topic should influence the similarity factor then Ac could be multiplied by the reviewer's level of competence, i.e.:

$$(4) \quad A_c = w_{r_j}(kw_c)(1 - (w_{p_i}(kw_c) - w_{r_j}(kw_c))) \ .$$

In general calculating Ac as in (3) is preferred than using formula (4) because it is arguable whether it is worthy to assign highly competent experts in specific topic to papers that are just a little bit related to this topic.

## 2.4. Describing Papers and Reviewers' Competences by a List of Conference Topics and Bidding

The combination of the two previously discussed methods aims to gather their advantages and if not to eliminate, then to significantly reduce the effect of their disadvantages.

Obviously the most reasonable combination is the following:

1. Authors and reviewers describe their papers, respectively competences, by using a list of conference topics (with multiple selection).

2. Then the software finds, for every single reviewer, a small amount of papers (let's say 20) that will be most interesting to him and suggest him to rate just this subset of papers, not all of the submitted ones.

3. Reviewers rate the papers suggested to them.

4. Optionally an Iterative Rating Method [5] can be used to predict any missing ratings.

5. If paper $p_i$ has been explicitly rated by reviewer $r_j$ then the weight of the corresponding edge $w(e_{p_i r_j})$ depends on the reviewer's rating only. However, if paper $p_i$ has not been rated by reviewer $r_j$ then the weight $w(e_{pirj})$ is calculated by using Jaccard's or Dice's formula over the lists of keywords describing $p_i$ and $r_j$. If a paper has not been rated by any reviewer and step 4 is omitted then the assignment algorithm can still find suitable reviewers to evaluate it. In this case the conference topics are used to calculate the similarity factors between the paper and all of the reviewers. If the paper has been rated the accuracy of the assignment will be probably higher, but even if not rated the assignment will not be random, but meaningful and accurate.

Although this looks to be the most appropriate combination of methods and sequence of use, none of the existing systems follows this pattern.

The MyReview System [17], for example, uses the conference topics to group reviewers in clusters then suggests a small amount of papers, to be rated, to every reviewer. The PC chair can use the Iterative Rating Method [5] in one or several iterations in order to predict the missing ratings. The problem of the implementation of The MyReview System is that it automatically assigns a neutral value to the edges (ratings) that have not been explicitly set by the reviewers or that have not been predicted by the IRM. I.e. conference topics are not directly used in calculating the similarity factors. It is better if a paper has not been rated by a reviewer then the weight of the corresponding edge to be calculated by using formula (1), (2) or (3) according to the conference topics selected by the author and the reviewer. Assigning a neutral value is highly close to assigning at random.

OpenConf [18], EasyChair [12], ConfTool [14] and Confious [15] also rely on both topics and bidding, however they do not suggest a small subset of papers to be rated by the reviewers, but reviewers are required to bid among all submitted papers. That is not convenient and never happens in case of a large number of submitted papers. ConfTool and Confious allow reviewers to filter the list of papers by topic, but just by a single topic. That helps, yes, but not much. If a paper has not been explicitly rated by a reviewer then all these systems rely on conference topics.

EasyChair do not use Jaccard's index or other complex distance measure, but just counts the number of keywords in common. If a paper has more than one common topic with the PC member, it will be regarded as if he chooses „I want to review this paper". If a paper has exactly one common topic with the PC member, it will be regarded as „I can review it" [12]. This, of course, is much less accurate than calculating the similarity factor by using Jaccard's formula. Here we have just 3 degrees of accuracy – 0; 0.5; 1, while Jaccard's formula allows smoother (continuous-like) change form 0 to 1.

It is not clear how exactly OpenConf, ConfTool and Confious calculate the similarity factors. There is no documentation on this issue at all.

CyberChair [13] also relies on both topics and bidding. Reviewers browse papers by topic and bids on the ones they want. If a paper has not been rated by a reviewer then the conference topics are taken into account. The unique feature of CyberChair is that it allows reviewers not only to state which topics they are competent in, but also to indicate how much exactly they are competent. This „bidding on topics"-like feature increases the accuracy of the assignment as the highly competent reviewers will be distinguished from those just a little bit competent.

Microsoft CMT [16] uses a greedy algorithm, assigning a paper to the reviewers who give the highest preference, but limiting the number of papers assigned to a reviewer to a threshold. When the system cannot find a reviewer, a matching of reviewers and paper topics is used [5]. Using a greedy algorithm as an assignment one does not guarantee high accuracy and even may leave papers without reviewers. More on this is discussed in part 3 – Influence of the assignment algorithm on the assignment accuracy.

The commence conference management system [19] relies just on bidding. No keywords are used to assist the assignment process.

An interesting solution of the assignment problem is provided by the GRAPE system [6]. It is a rule based expert system written in CLIPS. It relies on both conference topics and bidding, taking both into account when doing the assignment. However its fundamental assumption is to prefer topics matching approach over the reviewers' bidding one, based on the idea that they give assignments more reliability [6]. Reviewers' references are used just to tune the assignments. This system is interesting because it is the rule based. The assignment algorithm just navigates through the rules and executes them. Hence it is very easy to add new rules in order to change/improve its behavior, and it is possible to describe background knowledge, such as further constraints or conflicts, in a natural way [6].

### 2.5. Implicit Methods of Describing Papers and Competences

The implicit methods of describing papers and reviewers' competences use intelligent methods of extracting the needed description from the corresponding papers and also from online digital libraries and indexes such as DBLP [20], ACM Digital Library [21], CiteSeer [22], Google Scholar [23], Ceur WS [24] and others. These methods do not require users to select topics, as they are automatically extracted from the text.

Andreas Pesenhofer et al. [1] suggest that *the interest of a reviewer can be identified based on his/her previous publications available on the Internet*. For that reason the reviewer's first and last names are used to formulate a search query sent to CiteSeer and Google Scholar. The re-turned results construct the so-called „reviewer's profile". It contains all publications authored by the specified reviewer. Then Pesenhofer et al. compute the Euclidian distance between every paper submitted to the conference and every reviewer's publication based on the full-text indexed feature vector [1]. As a PC member has normally more than one publication in his profile, they keep only the smallest distance from all his documents to one submission [1]. This proposal seems to be computationally expensive as the scientists who usually act as reviewers have a very large number of publications, so calculating the distance between every paper and every single publication of every reviewer could be a very time/resource consuming task.

Stefano Ferilli et al. [7] propose a similar solution where *paper topics are extracted from its title and abstract, and expertise of the reviewers from the titles of their publications available on the Internet*. The proposed method assumes there is a predefined set of conference topics and *it tells which topics exactly apply to which papers/reviewers* and in this sense it eliminates the need of explicitly selecting topics by authors and reviewers. The title and the abstract of the submitted papers are analyzed and the words contained therein are stemmed according to the Porter's technique [8]. A Latent Semantic Indexing (LSI) [25] is applied to the set of all word stems in order to index the whole set of documents [7]. Then the titles of the reviewers' publications are obtained from DBLP and the same procedure is applied to them as well. To find out which topics apply to which papers/reviewers a number of queries, each corresponding to one conference topic, are performed on both papers and reviewers in the database previously indexed [7]. The proposal was evaluated on a real-world dataset built by using data from IEA/AIE 2005 conference. Evaluated by the conference organizers, the proposed method showed 79% accuracy on average. As to reviewers the resulting accuracy was 65% [7]. The results are valid in comparison with the manual assignment, so it is difficult to say how accurate this method substitutes the explicit selecting of conference topics done by authors and reviewers.

Another way of determining the most suitable reviewers for papers is the algorithm suggested by Marko Rodriguez and Johan Bollen [4]. Their approach is based on the premise that *a manuscript's subject domain can be represented by the authors of its references*. In comparison with the previous two methods this one does not exploit the titles and the abstracts, but the names of the authors who appear in the reference section. An important feature of this approach is that *it does not limit the set of potential reviewers to a predefined set of PC members*. It allows a conference management system to dynamically find reviewers on the Internet and invite them to register and evaluate the corresponding paper. The approach uses a co-authorship network, built from the authors in the reference section of a paper, their co-authors, the co-authors of the co-authors etc., and then a relative-rank particle-swarm algorithm is run for finding the most appropriate experts to be reviewers of the paper. The co-authorship network is defined by a graph composed of nodes that represent authors and edges that represent joint publication between two authors. Every edge has a weight representing the strength of tie between any two collaborating authors [4]. Rodriguez and Bollen use data from DBLP to build the co-authorship network. Initially the network is built from the names in the reference section of the specified paper. Then the co-authors of the authors already in the network are extracted from DBLP and added to the co-authorship network and so on. Although the proposed algorithm has a linear computational complexity in respect to the particle population, due to the large number of particles and the large number of steps needed for particle propagation the running time is far from small. The authors report when implemented in Java and run on Intel Core Duo it takes 1.674 seconds per article. Then for 200 papers the algorithm will need about 335 seconds. This is acceptable, but for offline execution only as most of the shared hosting companies will not allow a single script to run for so much time. If the application is hosted on an own server then running time is not an issue at all. As for the accuracy of the assignment the authors perform a pretty detailed analysis that proves their approach is really working and produce good results in respect to bidding.

The implicit methods of describing papers and competences use intelligent techniques to extract the needed descriptive data decreasing in this way the level of interactivity and saving time to both authors and reviewers. In this way of thinking they look to be promising tools in the modern dynamic and always in a hurry world we live. However it is arguably if and how correctly they can substitute the subjective human judgments. The implicit methods rely on external data sources on the Internet that are more or less inertial. For example, from authors' own experience, the table of content of conference proceedings is published in DBLP with months delay that can even grow to a year. So a system that uses DBLP as an external data source can not actually guarantee it uses up to date information that in case of highly dynamic rapidly changing sciences could be e big problem. If a paper references recently published papers, then the possibility of not finding these papers on DBLP is very high, increasing in this way the impossibility of constructing a relevant co-authorship network. Due to the large number of conferences Google Scholar also needs a lot of time for indexing papers. Similarly, CiteSeer and ACM DL contain sparse information. If a paper is not cited then it will never appear in CiteSeer. If it is not published in an ACM-supported conference it will probably never appear in the ACM DL as well.

Probably because of these reasons all commercially available conference management systems use explicit methods where the assignment does not depend on external data sources making it from some points of view more reliable.

## 3. Influence of the Assignment Algorithm on the Assignment's Accuracy

In some cases the assignment algorithm could not be evaluated separately from the data model and method of describing papers and competences. This is typical for implicit methods of describing as they actually do not use external data models for describing objects. The explicit methods, however, commonly result in building a similarity matrix (*figure 2*). It contains the weights of all edges of the bipartite graph used to represent the sets of papers and reviewers (*figure 1*). The way these weights are calculated is determined by the method of describing papers and competences. Then an assignment algorithm is run over the similarity matrix to suggest the reviewer(s) to be assigned to each paper. One and the same matrix can be used as an input of many different algorithms that will eventually produce different assignments (outputs). As all algorithms share the same input data then the accuracy of an algorithm can be measured by the total average weight of the assignments it suggests. Said in more formal words the algorithm's accuracy is measured by the weight of matching* (5). The higher the weight of matching the higher is the accuracy.

$$(5) \quad w(M) = \sum_{e \in M} w(e)$$

where w(M) – the weight of matching M;
w(e) – the weight of an edge e (i.e. the similarity factor between the paper and the reviewer connected with e) that belongs to M.

\* In the graph theory matching is a set of edges that do not share common vertices. In the context of conference management systems we use the same term matching to refer to the set of all assignments, although this violates (to some extent) the strict definition of matching as a single paper is evaluated by many reviewers and a single reviewer can evaluates many papers. However traditional matching algorithms known from graph theory are fully applicable here as they are run in many passes and on each pass they match one paper to just one reviewer, following in this sense the strict matching definition.

As equation (5) shows the accuracy of an assignment algorithm could be evaluated by the weight of matching. Then the weight is compared to an *etalon value* corresponding to the highest possible accuracy that could be achieved with the same dataset. Obviously the highest possible accuracy requires the matching to have a maximum weight. A number of advanced matching algorithms that do really guarantee finding the maximum-weighted matching have been proposed in the literature. The most commonly used and one of the most efficient is the algorithm of Kuhn and Munkres, also known as the Hungarian algorithm. Its optimized version runs in $O(n^3)$.

The MyReview System [17] is the only one commercially available system that reveals the exact assignment algorithm it uses – a modified version of the Kuhn and Munkres algorithm implemented by Prof. Miki Hermann [5]. Unfortunately the computational complexity of $O(n^3)$ of the Hungarian algorithm is not satisfactory when implemented in PHP, so the authors of The MyReview System suggest (in their official documentation [17]) that in case of more than 200 submitted papers the C-based implementation of the algorithm (distributed together with the system) should be used. Their experiments show that the PHP implementation running over 200 papers and 150 reviewers takes about 300 seconds to complete the assignment. The biggest problem however is that according to users experience in case of a large number of submitted papers the PHP implementation may not terminate [17] or cause a „not responding" behavior of the HTTP server. Of course the compiled code of the C implementation is many times faster, but it is distributed as a source code that has to be compiled first by using the right compiler by a competent programmer.

The exact assignment algorithms of the other systems are more or less unknown. EasyChair, for example, says it uses a special-purpose randomized algorithm [12] but no specific details are provided. It is written however that the algorithm needs plenty of time and runs in multiple passes each of which is about 1-2 minutes. The EasyChair's developpers recommend PC chairs to repeat the automatic assignment several times until the algorithm cannot improve the quality 3-4 attempts in a row [12].

As described in [5] the Microsoft CMT uses a greedy algorithm to perform the automatic assignment. As known, greedy algorithms use partial data for finding the optimal solution locally. A decision made by the algorithm at a given stage may be optimal for that stage, but to lead to a non-optimal global solution. In the context of the conference management system, using a greedy algorithm for automatic assignment of reviewers to papers may lead to the following problem: Imagine that $r_j$ is competent to review more papers than the threshold allows. Let $r_j$ is the only one competent to review paper $p_i$. But in case of sequential assignment (the algorithm first assigns all reviewers to paper i, then processes paper i+1) with a greedy algorithm, when the time comes for $p_i$ to be assigned, it is pretty possible that $r_j$ might be busy, i.e. already to have enough papers to review. If such a scenario happens $p_i$ will be left without a reviewer.

Paper [9] proposes a heuristic algorithm that overcomes the above mentioned problem by taking into account not only the similarity factors, but the number of reviewers competent to review a specified paper as well, and finally by assigning reviewers „in parallel". It works as follows: all columns of the similarity matrix, where papers correspond to columns and reviewers to rows, are sorted by the similarity factors in descending order, so that the first row of the matrix contains the most suitable reviewers for all papers, the second row contains the second most suitable reviewers and etc (*figure 7*). Then the matrix is analyzed and some of the similarity factors are modified with a correction that depends on the number of reviewers competent to review the specified papers. If this number seems not to be enough then the corresponding paper should be treated with higher priority. The algorithm iteratively modifies some of the similarity factors, deletes others if necessary and finally it assigns a reviewer to each one of the papers „in parallel". The phrase „in parallel" is not used in a sense of a parallel data processing, but to describe that the algorithm does not do any assignments until it finds suitable reviewers for all papers. Once it finds the most suitable and *feasible* reviewers for all papers then it actually does all the assignments simultaneously. In this way the algorithm guarantees that if there is at least one reviewer competent to review a specified paper, then the paper will have a reviewer assigned to it and the „already busy reviewer" problem can not happen.

as it has a direct impact on the quality and the authors' trust in the specified event. The manual assignment is feasible only in case of a small number of papers and reviewers. If the number of submitted papers and registered reviewers gets bigger and the PC chairs do not know their competences in details then the manual assignment is practically useless. In this case the only reasonable way of assigning is the automatic assignment as it can handle a large number of constraints simultaneously. The accuracy of the automatic assignment depends on both the method of describing papers and competences, and on the accuracy of the assignment algorithm. The explicit methods (see section 2 and 2.1 to 2.4) require that users explicitly provide additional information to describe their papers and competences. In contrast the implicit methods (refer to section 2 and 2.5) do not require any explicitly user provided information, but automatically fetch it from the submitted papers and the Internet.

Based on the assumption that nothing is more accurate than the one specified by the user himself, all commercially available conference management systems rely on the explicit methods of describing papers and competences. Their decision is further supported by the fact implicit methods usually rely on external data sources that are more or less inertial, that could be a big problem for rapidly chang-

| P1 | P2 | P3 | P4 | P5 |
|---|---|---|---|---|
| R1 => 0.40 | R1 => 0.54 | R5 => 0.27 | R1 => 0.27 | R1 => 0.25 |
| R5 => 0.38 | R5 => 0.42 | R4 => 0.25 | R5 => 0.22 | R4 => 0.14 |
| R2 => 0.20 | R4 => 0.30 | R1 => 0.21 | R4 => 0.17 | R5 => 0.09 |
| R3 => 0.18 | R3 => 0.20 | R2 => 0.14 | - | - |
| R4 => 0.17 | R2 => 0.10 | R3 => 0.13 | - | - |

**Figure 7.** A sample similarity matrix after sorting the columns by similarity factors in descending order. The „-" symbol means the corresponding similarity factor (weight) is 0 and will not be processed.

The computational complexity of this algorithm is $O(n^2)$ in the worst case scenario [10] while achieving an accuracy of about 98% (observed in experiments but not statistically proven yet) of the one derived with the maximum-weighted matching algorithm of Kuhn and Munkres. Moving to quadratic from cubic complexity allows the PHP implementation of the algorithm to complete the assignment in a reasonable time, commensurable to the typical response time of most web applications, even when processing a large number of submitted papers.

## 4. Conclusion

The process of assigning papers to reviewers is probably the most important activity in organizing conferences ing sciences. Currently the interest to the implicit methods is scientific rather than commercial, but some of them look very promising. The *table* summarizes and compares the main characteristics of the methods used by the existing conference management systems.

Obviously the most accurate of the existing explicit ways of describing papers and competences is the combination of both list of conference topics and bidding. According to the chosen topics the conference management system suggest a small subset of papers to be rated by each one of the reviewers. Then if a paper has been explicitly rated by a reviewer the assigned weight is used as a similarity factor, otherwise the similarity factor is calculated by taking into account the conference topics only, but not at random.

**Comparison of the existing methods of describing papers and competences**

|  | Accuracy * | Calculating the similarity factors ** | Advantages | Disadvantages |
|---|---|---|---|---|
| **Bidding only** (refer to section 2.1) | low-medium | Explicitly stated by reviewers | The similarity factors (although their small number) are 100% accurate. | Does not actually describe papers and competences, but their relationships (if stated). If a reviewer does not rate a paper there is no way to calculate how much he/she is competent to review it. Large number of random assignments. |
| **Bidding + IRM** (section 2.2) | medium-high | Explicitly stated + predicted by IRM and collaborative filtering | Predicts some of the missing preferences (similarity factors). | Increased level of interactivity. Reviewers are required to rate papers multiple times. Still – if a paper has not been rated by any reviewer, then the latter is assigned to it at random. |
| **Keywords/ conf. topics** (section 2.3) | Medium-high | Jaccard's or Dice's similarity | Describes both papers and competences independently of each other. A similarity factor could be accurately calculated for every pair (paper, reviewer). Low number of random assignments. | There is no theoretical guarantee that the similarity factors are 100% accurate, although in practice they are usually very accurate. |
| **Topics + bidding** (section 2.4) | high | Jaccard's or Dice's similarity + explicitly stated | Combines the advantages of the other explicit methods and reduces significantly their disadvantages. | - |
| **Implicit methods of describing** (section 2.5) | Not enough evidences to claim anything for sure, but probably medium-high | Depending on the assignment algorithm and the data model | Save time as they do not require any explicit user actions. No additional descriptions like "list of conference topics" are necessary. | Rely on external data sources on the Internet that are very inertial. It is not enough analyzed how correctly these intelligent methods could determine the papers' subject domain or the reviewers' competences. Slow/Computationally expensive. |

* The accuracy stated in *table* is based on personal opinion and judgments of the authors of this article.
** The conference management systems using a list of conference topics to describe papers and competences do not reveal the way of calculating the similarity factors. According to the authors of this article the Jaccard's index (1), its modified form (3) and the Dice's coefficient (2) are the most relevant ways of doing so.

In case of a large number of submitted papers and registered reviewers the automatic assignment is for sure more accurate than the manual one. However its accuracy strongly depends on both the method of describing papers and competences and the assignment algorithm. So any potential web architect designing a conference management system should pay a special attention to these two as they are crucial for the success of the event being organized.

## Acknowledgement

This paper is financed by project: Creative Development Support of Doctoral Students, Post-Doctoral and Young Researches in the Field of Computer Science, BG 051PO001-3.3.04/13, European Social Fund 2007-2013, Operational Programme „Human Resources Development".

*Yordan Kalmukov* received his BSc and MSc degrees in Computer Technologies at the University of Ruse in 2005 and 2006 respectively. In 2002 he has studied Applied Computer Technologies at Liverpool John Moores University as an Erasmus student funded by the EU.
Currently he is a PhD student at the Department of Computing, University of Ruse and teaches BSc students in web programming.
His research interests are: automated document management and categorization systems, web-based information systems, algorithms and data structures, and databases.

Contacts: JKalmukov@gmail.com

**Boris Rachev,** Assoc. Prof. Dr., Technical University of Varna, Department of Computer Sciences and Technologies. Area of scientific interests: databases, information systems and technologies, image databases, multimedia systems and technologies, computer graphics.

Contacts:
Technical University of Varna,
Dept. of Comp. Sciences and Technologies,
1 Studentska str., 9010 Varna
tel: (052) 383 407
e-mail: Bob_Ra@acm.org